\documentstyle[epsfig]{mn}

\def\gs{\mathrel{\raise0.35ex\hbox{$\scriptstyle >$}\kern-0.6em
\lower0.40ex\hbox{{$\scriptstyle \sim$}}}}
\def\ls{\mathrel{\raise0.35ex\hbox{$\scriptstyle <$}\kern-0.6em
\lower0.40ex\hbox{{$\scriptstyle \sim$}}}}

\begin{document}

\title 
[Submillimetre--radio redshift indicators] 
{Dust temperature and the submillimetre--radio 
flux density ratio as a redshift indicator for distant galaxies
}
\author
[A.\,W. Blain]
{A.\,W. Blain \\
Cavendish Laboratory, Madingley Road, Cambridge, CB3 0HE.
}
\maketitle 

\begin{abstract} 
It is difficult to identify the distant galaxies selected in 
existing submillimetre-wave 
surveys, because their positions are known at best to only several arcseconds.  
Centimetre-wave VLA observations are required in order to determine
positions to subarcsecond accuracy, and so to allow reliable optical 
identifications to be made. Carilli \& Yun pointed out that the ratio of the 
radio to submillimetre-wave flux densities provides a redshift indicator for 
dusty star-forming galaxies, when compared with the tight correlation between 
the far-infrared and radio flux densities observed in low-redshift galaxies. 
This method does provide a 
useful, albeit imprecise, indication of the distance to a submillimetre-selected 
galaxy. Unfortunately, 
it does not provide an unequivocal redshift estimate, 
as the degeneracy between the effects of increasing the redshift of a galaxy 
and decreasing its dust temperature is not broken.
\end{abstract} 


\section{Introduction} 

The intensity of synchrotron radio emission from shock-heated electrons in 
star-forming galaxies is known to be correlated tightly with their far-infrared 
emission from dust grains heated by the interstellar radiation field 
(see the review by Condon 1992). 
This far-infrared--radio correlation arises because both  
radiation processes are connected with 
ongoing high-mass star formation activity in a galaxy. 
The correlation is described 
accurately by the equation, 
\begin{equation} 
S_{\rm 1.4GHz} = [(1.7^{+1.0}_{-0.6}) \times 10^{-3}] \> 
( 2.58 S_{\rm 60{\mu}m} + S_{\rm 100{\mu}m} ),  
\end{equation} 
which links the flux densities of a galaxy in the 60- and 100-$\mu$m {\it IRAS} 
passbands and at a frequency of 1.4\,GHz in the radio waveband, 
$S_{\rm 60{\mu}m}$, $S_{100\rm{{\mu}m}}$ and $S_{\rm 1.4GHz}$, respectively.

A reasonable pair of template spectral energy distributions (SEDs), which 
describe dusty galaxies at the relevant frequencies in equation (1), are shown in 
Fig.\,1. The observed flux densities of the luminous dusty galaxy Arp\,220 are 
also shown. The model SED derived in the far-infrared
waveband by Blain et al.\ (1999b), based on {\it IRAS} and {\it ISO} galaxy counts, 
is continued into the radio waveband directly 
using equation (1). The SED and galaxy evolution models from that paper  
can be combined with the far-infrared--radio correlation 
(equation 1) to predict the counts of faint radio galaxies. The predicted 
count of galaxies brighter than 10\,$\mu$Jy at a wavelength of 
8.4\,GHz is 0.8\,arcmin$^{-2}$, 
in excellent agreement with the observed value of $1.0 \pm 0.1$\,arcmin$^{-2}$
(Partridge et al. 1997). 

The slope of the SED changes abruptly at a wavelength of about 
3\,mm, at which the dominant contribution to the SED changes from 
synchrotron radio 
emission to thermal dust radiation. Because the far-infrared--radio 
correlation links the flux densities on either side of this spectral break, it 
could be exploited to indicate the redshift of the galaxy (Carilli \& Yun 1999).
Carilli \& Yun  
calculated that the radio--submillimetre flux density ratio of a distant dusty 
galaxy, which lies on the far-infrared--radio correlation, should be
\begin{equation} 
{ {S_{\rm 1.4GHz}} \over { S_{\rm 850{\mu}m} } } = 3.763
(1+z)^{1.007(\alpha_{\rm radio} - \alpha_{\rm submm})},
\end{equation} 
as a function of redshift $z$. $\alpha_{\rm radio}$ and $\alpha_{\rm submm}$ are 
the spectral indices of the SED, $S_\nu \propto \nu^\alpha$, in the radio and 
submillimetre wavebands, respectively. Typically, 
$\alpha_{\rm radio} \simeq -0.8$ and $\alpha_{\rm submm} \simeq 3.0$ to 3.5. 
The submillimetre-wave spectral index $\alpha_{\rm submm}$ is the sum of 
the Rayleigh-Jeans spectral index ($\alpha=2$) and $\beta$, the spectral index 
in the dust emissivity function $\epsilon_\nu \propto \nu^\beta$. Carilli \& 
Yun also derived empirical flux density ratio--redshift relations from 
the SEDs of Arp\,220 and M82. The  
spread of the redshift values that correspond to a fixed flux density ratio 
across their 
four models corresponds to an 
uncertainty in the predicted redshift of about $\pm 0.5$. 
 
The sensitive SCUBA submillimetre-wave camera at the JCMT (Holland et al.\ 
1999) has been used to detect high-redshift dusty galaxies at a wavelength of 
850\,$\mu$m (see Barger, Cowie \& Sanders 1999a and references within). 
These galaxies are probably the high-redshift counterparts to the low-redshift 
ultraluminous infrared galaxies, and so their SEDs might be expected to follow 
the far-infrared--radio correlation. In this case, the ratios of their radio 
and submillimetre-wave flux 
densities could be substituted into equation (2) to indicate their 
redshifts. 

The observed radio and submillimetre-wave flux densities of the 
submillimetre-selected galaxies SMM\,J02399$-$0136 
(at $z=2.808 \pm 0.002$;  
Frayer et al.\ 1998; Ivison et al.\ 1998) and SMM\,J14011+0252 (at 
$z=2.5653\pm 0.0003$; Frayer et al.\ 1999; Ivison et al. submitted),
the $z=4.69$ QSO BR\,1202$-$0725 (Isaak et al. 1994; Kawabe et al.\ 1999), 
the $z=2.29$ galaxy 
{\it IRAS} F10214+4724 (Rowan-Robinson et al.\ 1993) and the 
$z=0.02$ {\it IRAS} galaxy Arp\,220 can be substituted into equation (2) to 
obtain predictions for their redshifts. If $\alpha_{\rm submm} = 3.0$, then 
these are $z=2.9 \pm 0.3$, 
$3.8 \pm 0.4$, $3.9 \pm 0.5$, $2.8 \pm 0.2$ and $1.1 \pm 0.1$ 
respectively; if $\alpha_{\rm submm} = 3.5$, then 
$z=2.3 \pm 0.3$, $3.3 \pm 0.4$, $3.1 \pm 0.4$,  
$2.2 \pm 0.3$ and $1.0 \pm 0.1$ respectively. 
The technique provides a coarse indication of the redshift 
of these five galaxies, which are all known to lie at redshifts 
within the bounds of the range of predictions made by Carilli \& Yun's 
four models. For an application of this method to the 
radio and submillimetre-wave fluxes of galaxies detected in the 
SCUBA lens survey 
see Smail et al.\ (1999).  
What are the systematic effects that limit the reliability 
of the inferred redshifts?

\begin{figure}
\begin{center}
\epsfig{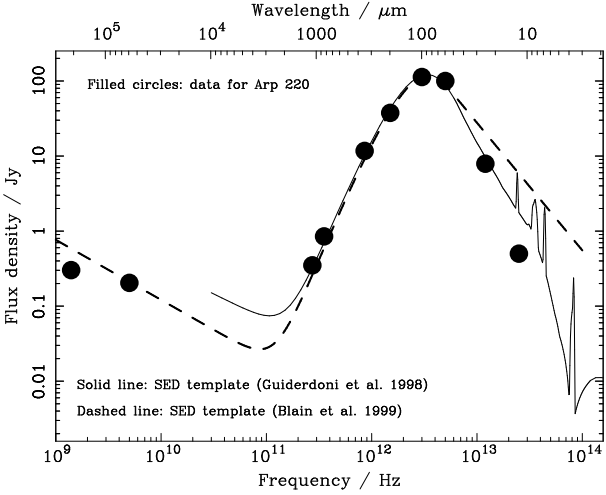}
\end{center}
\caption{ 
The template SEDs used to describe dusty galaxies by Guiderdoni et al.\ 
(1998; solid line) and Blain et al.\ (1999b; dashed line). The dashed line is 
normalized to match the far-infrared--radio correlation described in equation 
(1) (Condon 1992). At wavelengths longer than about 3\,mm, the SED is 
dominated by synchrotron radio emission with a spectral index 
$\alpha_{\rm radio} \simeq -0.8$. At shorter wavelengths the SED is 
dominated by the thermal radiation 
from dust grains with an emissivity index $\beta = 1.5$ and a temperature 
$T=38$\,K. The observed SED of Arp 220 (filled circles) 
is also shown. Guiderdoni et al.'s template includes mid-infrared spectral 
features (see Xu et al.\ 1998). 
} 
\end{figure}  

\section{The far-infrared--radio correlation at high redshifts} 

The far-infrared--radio correlation is based on observations 
of a range of low-redshift galaxies: low- and high-luminosity spiral galaxies, 
irregular star-forming dwarf galaxies and more luminous infrared galaxies, 
with luminosities up to about $5 \times 10^{11} h^{-2}$\,L$_\odot$. One 
factor that could modify the general properties of the SEDs of dusty 
galaxies at 
high redshifts is the increase in the temperature of the cosmic microwave 
background radiation (CMBR) from its value $T_{\rm CMBR} = 2.726$\,K at $z=0$, 
as $(1+z)$. If a certain type of galaxy contains dust at a temperature $T_0$ at 
$z=0$, then the dust temperature of an identical galaxy at redshift $z$ would 
be approximately, 
\begin{equation} 
T(z) = \left\{ T_0^{4+\beta} + T_{\rm CMBR}^{4+\beta} 
\left[ (1+z)^{4+\beta} - 1 \right] \right\}^{ 1 / (4 + \beta) }. 
\end{equation} 
The bolometric far-infrared luminosity of the galaxy is not changed, and so 
the SED in the radio waveband should remain the same. 

The increasing temperature of the CMBR has one further effect. There is a 
redshift above which the energy density in the CMBR exceeds that 
in the magnetic field in the interstellar medium of the observed galaxy. 
Thus 
synchrotron radio emission will be suppressed due to the cooling of 
the relativistic electrons by the inverse Compton scattering of CMBR 
photons at all higher redshifts. Carilli \& Yun (1999) estimate 
that this will typically occur at redshifts $z \gs 6$ for ultraluminous 
infrared galaxies and at $z \gs 3$ for a high-redshift analogue of the 
Milky Way (Carilli, private communication). 
This effect is not included in 
the calculations carried out below; however, the potential deficit in the 
radio flux density from ultraluminous 
galaxies at $z \gs 6$ should be borne in mind.

The SED from a galaxy at any frequency in the radio, submillimetre 
and far-infrared wavebands 
can be found by adding the spectrum of thermal dust emission, $f_\nu(T)$, to 
the associated radio spectrum derived using equation (1). However, the 
relative normalization of the 
radio and far-infrared spectra depends on the dust 
temperature $T$ (see Fig.\,2). 
Because the far-infrared--radio correlation is defined at 
$z=0$, the increasing CMBR temperature at 
higher redshifts, discussed above, affects the SED in the 
far-infrared waveband but not in the radio 
waveband. The full expression for the flux density of a galaxy at redshift $z$ 
at an observing frequency $\nu$ is, 
\begin{eqnarray} 
\lefteqn{\nonumber
S_\nu \propto 
[(1.7^{+1.0}_{-0.6}) \times 10^{-3}]
\left[ 2.58 f_{5{\rm THz}}(T_0) + f_{3{\rm THz}}(T_0) \right] 
\times }\\
& & 
\>\>\>\>\>
\left[ { \nu(1+z) \over {1.4\,{\rm GHz}} } \right]^{\alpha_{\rm radio}}
+ f_{\nu(1+z)}[T(z)] \> 
{ { \int f_\nu'(T_0) \,{\rm d}\nu' } \over { \int f_\nu' [T(z)] \,{\rm d}\nu'} }.
\end{eqnarray}   
$f_{5{\rm THz}}$ and $f_{3{\rm THz}}$ are the flux densities obtained 
after 
integrating the SED $f_\nu$ 
across the 60- and 100-$\mu$m {\it IRAS} passbands and  
dividing by the bandwidth, respectively. 
Equation (4) can be used to predict the 1.4-GHz and 850-$\mu$m flux 
densities, which are required to investigate the 
reliability of the radio--submillimetre 
flux density ratio as a redshift indicator. 
The subtle modification introduced to
the ratio by the greater CMBR temperature at high redshifts is included. 
This correction is expected to be most significant for either cool galaxies with 
$T_0 \ls 20$\,K or very high-redshift galaxies at $z \gs 10$. 

\section{Uncertainties} 

Galaxies with a range of different dust temperatures have flux density ratios 
that lie on the far-infrared--radio correlation. This is because the 
far-infrared flux densities in the correlation (equation 1) are evaluated at 
60- and 100-$\mu$m, wavelengths close to the peak of the SED of a 
typical dusty galaxy. Hence, shifting the position of the peak of the SED, 
by changing the dust temperature $T$, makes little difference to the bolometric 
far-infrared flux density and thus to the far-infrared--radio flux density 
ratio. This weak temperature dependence is illustrated in Fig.\,2, in which the 
constant of proportionality in the far-infrared--radio correlation 
(equation 1) is shown as a function of dust temperature $T$. The 
radio flux density is assumed to be proportional
to the bolometric far-infrared luminosity, and the 
far-infrared SED is assumed to take the form shown in Fig.\,1, 
with $\beta=1.5$. 
The radio flux density is then 
divided by the far-infrared flux density 
in the 60- and 100-$\mu$m {\it IRAS} passbands to yield the constant 
of proportionality, which is assumed to match the standard value of 
$1.7 \times 10^{-3}$ exactly if $T=50$\,K. The horizontal dotted lines in 
Fig.\,2 show the magnitude of the observed scatter in the empirical 
far-infrared--radio correlation (Condon 1992). Hence, the dispersion 
in the relation expected owing to different dust temperatures throughout the 
wide range from 20 to 70\,K is only comparable with the intrinsic scatter in 
the correlation. 

\begin{figure}
\begin{center}
\epsfig{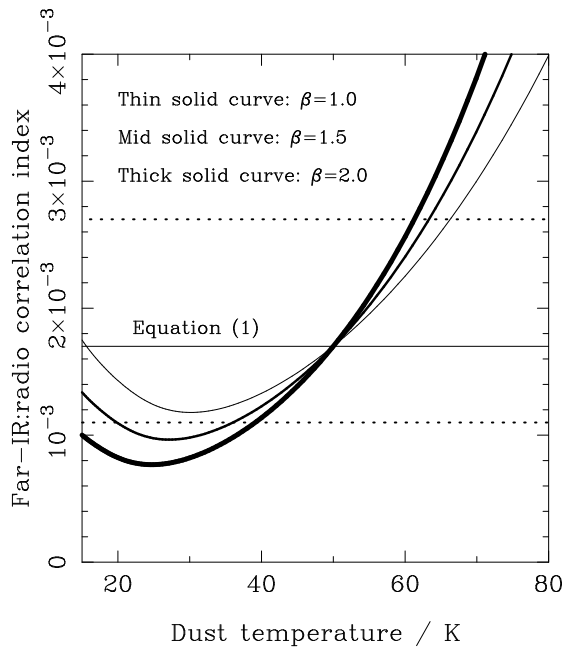}
\end{center}
\caption{The constant of proportionality predicted in the far-infrared--radio 
correlation (equation 1) as a function of dust temperature. The radio 
flux density is assumed to be proportional 
to the bolometric far-infrared luminosity. The 
far-infrared flux density in the {\it IRAS} passbands (equation 1) 
is obtained by integrating the SED 
over these passbands and adding the results in 
the appropriate proportion. 
The observed value of the constant 
and its uncertainty are shown by the horizontal solid and dotted lines 
respectively. The predictions are normalised to the observed relation 
at a temperature of 50\,K, which is typical of observed ultraluminous 
infrared galaxies.
}
\end{figure} 

However, the effect of modifying either the 
dust temperature $T$ or the emissivity index $\beta$ on the ratio of the 
flux densities of a star-forming galaxy in the submillimetre and radio 
wavebands is much greater, as 
shown by the five different model SEDs in Fig.\,3. Three different dust 
temperatures $T=20$, 40 and 60\,K are included, each with an emissivity index 
$\beta = 1.5$. In addition, the SED is calculated for the $T=40$\,K model with 
$\beta = 1.0$ and 2.0. 

The predicted forms of the submillimetre--radio flux density ratio as a function of 
redshift for all five of the template SEDs shown in Fig.\,3 are calculated using 
equation (4) and shown in Fig.\,4(a). The results of Carilli \& Yun (1999), 
calculated using equation (2), are also shown for comparison. 
Carilli \& Yun's equation provides a good description of the 
1.4-GHz:850-$\mu$m flux density ratio if the dust temperature at $z=0$, 
$T_0 \gs 60$\,K. However, the positions of the curves in Fig.\,4(a) can be 
quite 
different if the dust temperature $T_0 \ls 60$\,K. Hence, the redshift that 
would be assigned to a galaxy using the 1.4-GHz:850-$\mu$m flux density ratio 
alone, in the absence of knowledge of both the appropriate dust temperature 
and emissivity index, is very uncertain. This is true even if the far-infrared--radio 
correlation is assumed to be free from any intrinsic scatter, and any 
additional contribution to the radio flux density from an active galactic 
nucleus (AGN) is neglected. The scatter in the far-infrared--radio correlation 
is about 0.2\,dex, and because the power-law index of the lines in 
Fig.\,4(a) is about $-2$ at $z \sim 2$, an additional 0.1\,dex 
($\simeq 25$\,per cent) uncertainty would be expected. See Carilli \& Yun (1999) 
for a discussion of the effects of AGN.

It is interesting to replot the curves in Fig.\,4(a) as a function of the combined 
redshift--dust temperature parameter $(1+z)/T$. This is the quantity that is 
constrained by measuring the position of the peak of the thermal 
dust emission component of the SED by combining observations in the 
far- and mid-infrared wavebands (Blain 1999a). The results are shown in 
Fig.\,4(b). Because the radio flux density is produced by a non-thermal 
emission mechanism, a measurement of the submillimetre--radio flux density
ratio might be expected to break the degeneracy between temperature and 
redshift. However, because the differences between the curves in Fig.\,4(b) are 
not much greater than the scatter in the far-infrared--radio correlation, the 
degeneracy remains.

\subsection{Dust temperatures in luminous infrared galaxies} 

Uncertainties in the dust temperature of submillimetre-selected 
galaxies can lead to a significant uncertainty in the 
redshift that would be derived from the relations shown in Fig\,4(a). 
What is known about the dust temperatures of galaxies and quasars? 

In spiral galaxies, for example the Milky Way (Sodroski et al.\ 1997) 
and NGC\,891 (Alton et al.\ 1998; Israel, van der Werf \& Tilanus 1999), 
the observed SED can be adequately represented by 
a combination of different dust components with temperatures in the 
range from 15 to 30\,K. In the core of NGC\,891, Israel et al. (1999) 
report a compact component of warmer dust at $T \gs 50$\,K. 

Franceschini, Andreani \& 
Danese (1997) 
observed a sample of low-redshift 60-$\mu$m {\it IRAS} 
galaxies at a wavelength of 1.25\,mm, the luminosities of which were  
distributed between 10$^9$ and 10$^{11}$\,L$_\odot$. 
The ratio of the 60-$\mu$m and 
1.25-mm flux densities of these galaxies 
contains some information about the emissivity index and 
temperature of their interstellar dust. 
If only a single population of isothermal dust is 
present, then the scaling relation reported between the flux density ratio 
and the 
bolometric luminosity $L_{\rm bol}$ of the galaxies can be converted 
directly into a dust temperature--luminosity relation. If  
the emissivity index $\beta = 1.0$, then the 
dust temperature 
$T/{\rm K} \simeq 34 (L_{\rm bol} / 10^{10}\,{\rm L}_\odot)^{0.07}$, and if 
$\beta = 1.5$, then 
$T/{\rm K} \simeq 25 (L_{\rm bol} / 10^{10}\,{\rm L}_\odot)^{0.03}$. Hence, 
in this sample, dust temperatures appear to be relatively cool and there 
is only a weak dependence of temperature on luminosity. 
Note, however, that  
the 1.25-mm flux densities observed in this sample could be enhanced by 
the presence of either 
synchrotron/free--free radio emission or an 
energetically insignificant population of cool dust grains  in the 
observed galaxies, as can be seen from the shape of the  
SEDs shown in Fig.\,3. The dust temperature required to account for 
the data would be increased if any of these contributions were 
present.
An extensive survey being carried out by Dunne et al. 
(1999) using SCUBA at shorter wavelengths of 450 and
850\,$\mu$m will provide  
a more definitive result. 

\begin{figure}
\begin{center}
\epsfig{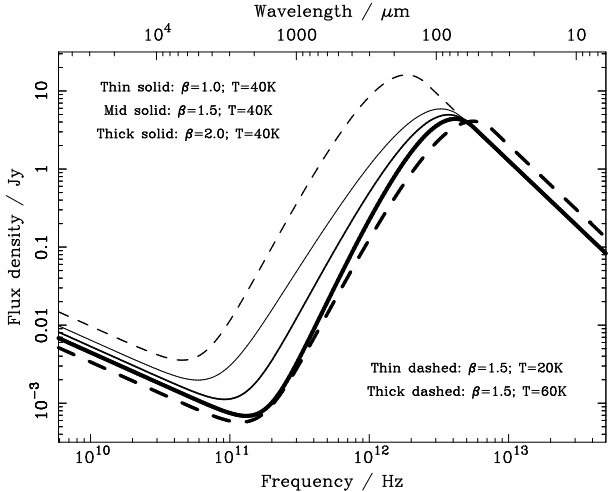}
\end{center}
\caption{Examples of SEDs defined by different values of the $z=0$ dust 
temperature $T_0$ and dust emissivity index $\beta$, all normalised to the same 
flux density at 60-$\mu$m. The SEDs are very different in the submillimetre 
waveband, but very similar at both far-infrared wavelengths
of 60 and 100\,$\mu$m and in the radio waveband.
The 1.4-GHz:850-$\mu$m flux density ratios predicted in all five models shown 
here are compared in Fig.\,4.
}
\end{figure}

For more luminous galaxies, temperatures fitted to the observed SEDs of 
low-redshift 
ultraluminous infrared galaxies, such as Arp\,220 and 
Mrk\,231, tend to lie close to $50 \pm 10$\,K. This temperature also 
provides a reasonable fit to the SED of the two submillimetre-selected 
galaxies with known redshifts, 
SMM\,J02399$-$0136 and SMM\,J14011+0252 (Ivison et al.\ 1998, 1999), and to 
the temperature of 
the coolest dust component of the high-redshift galaxies and quasars 
that were 
observed at a wavelength of 350-$\mu$m by Benford et al.\ (1999). For the 
three exotic gravitationally 
lensed high-redshift dusty galaxies and quasars selected in a 
variety of ways, IRAS\,F10214+4724, the Cloverleaf quasar 
H\,1413+117 and APM\,08279+5255, higher dust temperatures are typically 
found, in the range 75, 75 and 110\,K respectively (see the discussion 
of their SEDs by Blain 1999b). 

In order to provide consistent fits to the background radiation intensity 
and source counts in the far-infrared and submillimetre wavebands, assuming 
an isothermal population of high-redshift galaxies, both 
Blain et al. (1999b) and Trentham, Blain \& Goldader (1999) found that 
typical 
dust temperatures of order 40\,K were required.
If a range of dust temperatures are 
considered, then 40\,K is the likely luminosity-weighted average 
temperature. 

The results of fits to the SEDs of dusty galaxies tend to indicate that 
dust temperatures are of order 20\,K in low-luminosity spirals, of order 
40--50\,K in luminous {\it IRAS} galaxies, and perhaps higher in 
individual exotic high-redshift galaxies. At present, 
it seems reasonable to assume a temperature of order 40--50\,K for 
submillimetre-selected galaxies in the absence of other information. 
If a temperature of 40\,K is assumed, then the results of the Carilli \& 
Yun (1999) formula (equation 2) are in reasonable agreement with the 
results of the calculations presented here at least if $\alpha_{\rm submm} 
= 3.5$ is assumed in equation (2). 

A particular concern could 
be the potential misidentification of a submillimetre-selected galaxy 
with a low-redshift spiral galaxy, 
in which a dust temperature of about 20\,K would be expected, rather 
than with a true, hotter and more distant counterpart, which would be 
expected to have the same value of $T/(1+z)$. From Fig.\,4(a), it is clear 
that a 20-K galaxy at $z \simeq 0.5$ and a 40-K galaxy at $z \simeq 2$ 
produce the same ratio of the 1.4-GHz and 850-$\mu$m flux densities. 
A real example of this 
type of degeneracy is seen in the combined submillimetre, radio, optical 
and near-infrared data that was used by Smail et al.\ (1999) to associate 
a background extremely red object (ERO) with a submillimetre source 
in the field 
of the cluster 
Cl\,0939+47, rather than a $z=0.33$ dusty spiral galaxy.  

\begin{figure}
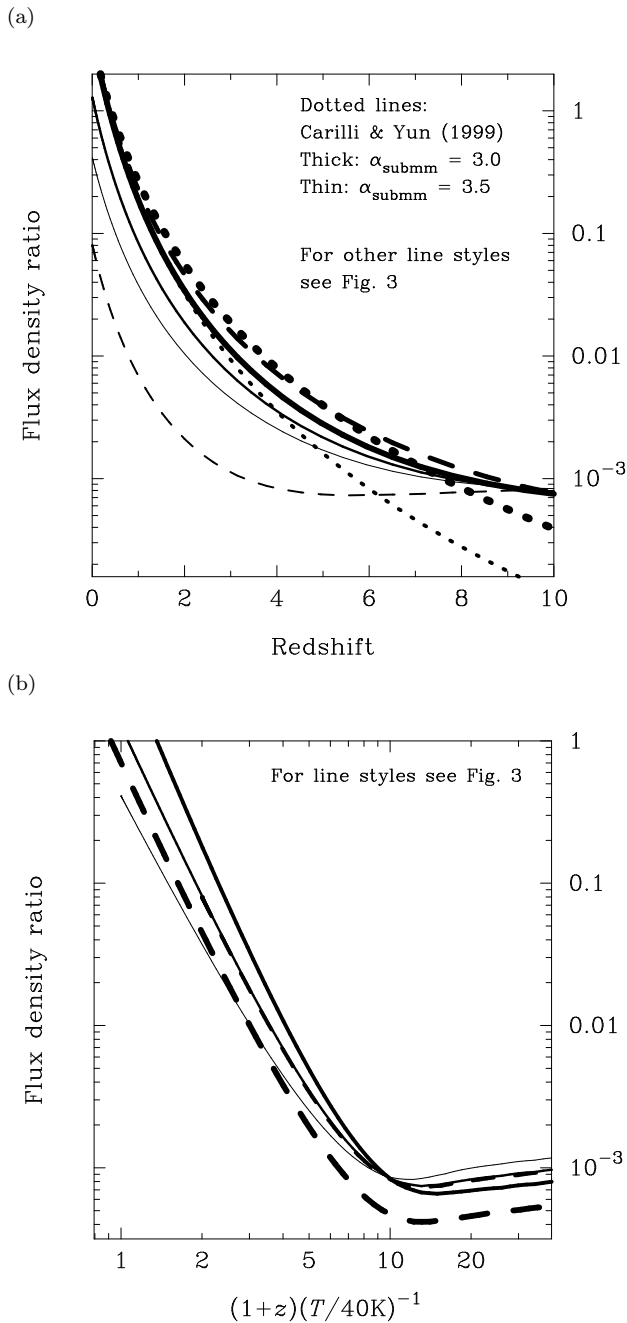
 
(a) 
\begin{center}
\epsfig{file=fig4a.ps, width=7.85cm, angle=-90} 
\end{center} 
(b)
\begin{center} 
\epsfig{file=fig4b.ps, width=7.85cm, angle=-90}
\end{center}
\caption{(a) The 1.4-GHz:850-$\mu$m flux density ratios expected for the five 
model SEDs shown in Fig.\,3 as a function of redshift $z$. A considerable 
range of redshifts could be deduced from a measurement of this ratio if the 
temperature $T$ or emissivity index $\beta$ of the dust in the galaxy being 
observed was uncertain. The curves from Carilli \& Yun (1999) are plotted 
assuming $\alpha_{\rm radio} = -0.8$. As stressed by Carilli \& Yun, if the 
1.4-GHz:850-$\mu$m flux density 
ratio is greater than 0.1, then the redshift $z$ is 
likely to be less than 1, and if the ratio is less than $10^{-2}$, then 
$z$ is likely to be greater than 2. 
(b) The same flux density ratio plotted as a function of 
$(1+z)/T$. In (b) the curves all overlap, indicating that it is difficult to break 
the degeneracy between the redshift $z$ and dust temperature $T$ of a galaxy 
using a measurement of the radio--submillimetre flux density ratio.}
\end{figure} 

\section{Other redshift indicators}

\subsection{Mid- and far-infrared photometry} 

The `redshifted dust temperature' of a galaxy, $T/(1+z)$, can be determined by 
measuring the frequency of the peak of the far-infrared dust component of
the SED (see Figs\,1 and 3). Because of the thermal spectrum, the redshift $z$ 
and the dust temperature $T$ cannot be determined independently. 
Locating the peak frequency requires both long- and short-wavelength 
observations. These could be obtained by combining ground-based 
millimetre-wave observations, for example using the second-generation 
1.1-mm bolometer camera BOLOCAM (Glenn et al.\ 1998), with far- and 
mid-infrared observations made at wavelengths of 24, 70 and 160\,$\mu$m 
using the MIPS instrument on the {\it SIRTF} satellite: for more 
details see Blain (1999a).

\subsection{Mid-infrared spectral features} 

Spectral features produced by emission from polycyclic aromatic 
hydrocarbons (PAHs) and atomic fine-structure lines in the restframe 
mid-infrared waveband at restframe wavelengths of order 10\,$\mu$m 
(see Fig.\,1) could be exploited to obtain photometric redshifts for distant 
galaxies (Xu et al.\ 1998). At shorter wavelengths, the prospects for 
obtaining photometric redshifts using the 3--10-$\mu$m {\it SIRTF} IRAC 
camera have been discussed recently by Simpson \& 
Eisenhardt (1999). Photometric redshifts deduced from  
these features will not be subject to the dust
temperature--redshift degeneracy. 

\subsection{Submillimetre--optical flux density ratios} 

The identification of optical counterparts to submillimetre-selected 
galaxies usually requires a radio observation and a great deal of observing 
time (see for example Ivison et al.\ 1998, 1999). 
The optical magnitudes and 
colours of heavily-obscured submillimetre-luminous galaxies are expected to 
extend over wide ranges, and so the derivation of redshift information 
from broad-band optical photometry will probably require careful 
individual analysis of each submillimetre-selected galaxy. The results of
photometric and spectroscopic observations of likely optical counterparts to
submillimetre-selected galaxies (Smail et al.\ 1998; Barger et al.\ 1999b) 
are consistent with ratios of 850-$\mu$m and optical $I$-band flux densities 
that are scattered by about an order of magnitude across the sample.

\section{Practical radio follow-up of future millimetre-wave surveys} 

In order to exploit the redshift information provided by joint radio and
submillimetre-wave observations, follow-up radio observations of 
submillimetre-selected galaxies must be acquired in a reasonable time.

In the next few years the most rapid survey for distant dusty galaxies 
will probably be made using the 1.1-mm BOLOCAM camera (Glenn et al.\ 1998). 
The detection rate of high-redshift galaxies at this wavelength 
is expected to be maximized at a survey depth of 10\,mJy (Blain 1999a), 
based on the latest deep 
counts of galaxies detected by SCUBA (Blain et al.\ 1999a).
At the Caltech Submillimeter Observatory (CSO), BOLOCAM should be 
able to map the sky to a 5$\sigma$ sensitivity of 10\,mJy at a rate of 
about 0.1\,deg$^2$\,hr$^{-1}$, and detect galaxies at a rate   
of about 4\,hr$^{-1}$. Fitted to the US--Mexican 50-m Large Millimeter 
Telescope, the mapping speed would be 1.4\,deg$^2$\,hr$^{-1}$ and the 
detection rate about 50\,hr$^{-1}$. 
For comparison, the 
mapping rate of SCUBA at 850\,$\mu$m to 
the same flux density limit is about 10$^{-5}$\,deg$^2$\,hr$^{-1}$, and the 
detection rate is about 0.2\,hr$^{-1}$ (Blain 1999a). 

In order to detect the radio emission from the 90\,per cent of
BOLOCAM-selected galaxies that are expected to lie at $z \ls 5$ (Blain et
al.\ 1999b), a 1.4-GHz VLA image that is about 300 times deeper in flux density 
than the BOLOCAM survey would be required. After the upgrade 
programme, the VLA will be able to reach the required 5$\sigma$ sensitivity of 
35\,$\mu$Jy in a 1-hr integration. The 30-arcmin primary beam of the 
VLA is expected to contain about 8 BOLOCAM detections. Hence, radio 
counterparts and thus temperature--redshift information can be obtained for 
the great majority of the galaxies detected in a millimetre-wave BOLOCAM/CSO 
survey, if about 15\,per cent of the total time involved in the survey is spent at the 
upgraded VLA. 

\section{Conclusions}

Radio observations of submillimetre-selected galaxies are crucial in order to 
make reliable optical identifications. The ratio of the radio and 
millimetre/submillimetre-wave flux densities also provides information about 
the redshift of the galaxy (Carilli \& Yun 1999). However, the 
temperature and emissivity of dust in distant galaxies has a 
particularly significant effect on the ratio of the ratio of the
submillimetre-wave and radio flux densities. For reasonable 
values of the dust emissivity and temperature, the 
radio--submillimetre flux density ratio cannot be used to break the 
degeneracy between the dust temperature and redshift of a distant 
dusty galaxy. 

\section*{Acknowledgements}
I thank Chris Carilli, Kate Isaak, Rob Ivison, Richard McMahon, Kate Quirk, 
Ian Smail, 
Min Yun and an anonymous referee 
for helpful comments. This research has
made use of the NASA/IPAC Extragalactic Database (NED) which is operated by
the Jet Propulsion Laboratory, California Institute of Technology, under
contract with  
the National Aeronautics and Space Administration.

\end{document}